\documentclass[iop]{emulateapj}
\usepackage{graphicx}
\usepackage{epsfig}
\usepackage{epstopdf}

\begin{document}

\shorttitle{Non-thermal Line Broadening of Lower Transition Region Lines}
\title{Why Is Non-thermal Line Broadening of Spectral Lines in The Lower Transition Region
  of the Sun Independent of Spatial Resolution?}
\author{B. De Pontieu\altaffilmark{1,2}}
\author{S. McIntosh\altaffilmark{3}}
\author{J. Martinez-Sykora\altaffilmark{4,1}}
\author{H. Peter\altaffilmark{5}}
\author{T.M.D. Pereira\altaffilmark{2}}


\affil{\altaffilmark{1}Lockheed Martin Solar \& Astrophysics Lab, Org.\ A021S,
  Bldg.\ 252, 3251 Hanover Street Palo Alto, CA~94304, USA}

\affil{\altaffilmark{2}Institute of Theoretical Astrophysics,
  University of Oslo,   %
  P.O. Box 1029 Blindern, N-0315 Oslo, Norway}

\affil{\altaffilmark{3}High Altitude Observatory, National Center for
  Atmospheric Research, P.O. Box 3000, Boulder, CO 80307}

\affil{\altaffilmark{4}Bay Area Environmental Research Institute,
  Sonoma, CA~94952, USA}

\affil{\altaffilmark{5}Max-Planck-Institut f\"ur
  Sonnensystemforschung, Justus-von-Liebig-Weg 3, D-37077 G\"{o}ttingen, Germany} 

\begin{abstract}
  Spectral observations of the solar transition region (TR) and corona
  show broadening of spectral lines beyond what is
  expected from thermal and instrumental broadening. The remaining
  non-thermal broadening is significant (5-30 km/s) and correlated
  with the intensity. Here we study spectra of the TR \ion{Si}{4} 1403 \AA\ 
  line obtained at high resolution with the Interface Region Imaging
  Spectrograph (IRIS). We find that the large improvement in spatial 
  resolution (0.33\arcsec) of IRIS compared to previous spectrographs
  (2\arcsec) does not resolve the non-thermal line broadening which
  in most regions remains at pre-IRIS levels of about 20 km/s. 
 This invariance to spatial resolution indicates that the processes behind the
 broadening occur along the line-of-sight
 (LOS) and/or on spatial scales (perpendicular to the LOS) smaller than 250 km. Both
  effects appear to play a role. 
  Comparison with IRIS chromospheric
  observations shows that, in regions where the LOS is more
  parallel to the field, magneto-acoustic shocks driven
  from below impact the TR and can lead to significant non-thermal line
  broadening. This scenario is
  supported by MHD simulations. While these do not show enough
    non-thermal line broadening, they do reproduce the
  long-known puzzling correlation between non-thermal line broadening
  and intensity. This correlation is caused by the shocks, but only if non-equilibrium ionization is taken into
  account. In regions where the LOS is more perpendicular to the field, the
  prevalence of small-scale twist is likely to play a significant role
  in explaining the invariance and correlation with intensity.
\end{abstract}

\keywords{waves \-- Sun:atmospheric motions \-- Sun:magnetic fields
  \-- Sun: chromosphere \-- Sun:transition region \-- Sun:corona}

\section{Introduction} 
\label{sec:intro}

The transition region between the chromosphere and corona has
long puzzled solar physicists \citep[see][and references therein]{Mariska1992}. Spectroscopic
observations of the first three moments of TR spectral lines have
presented difficult challenges to observers and theorists alike. For
example, low-resolution intensity measurements of TR lines have been
difficult to reproduce in coronal heating models leading to a debate
about the TR magnetic topology and the role of unresolved fine
structure (UFS) which recent IRIS observations
may have finally resolved \citep{Hansteen14}. Similarly,
low-resolution observations of TR velocities have revealed a pervasive redshift
\citep{Peter99,Peter99b} that is thought be a signature of the dominant heating mechanism in the
corona and which remains the subject of vigorous debate \citep{Peter2004,Hansteen10}. 
Finally, the non-thermal line broadening of TR lines has been interpreted as a signature of a 
variety of physical processes such as small-scale flows, coronal 
nanoflares, Alfv\'en waves, turbulence, etc. \citep{Mariska1992}. One potentially
important clue to the nature of non-thermal broadening
is the puzzling and unexplained correlation between the intensity and non-thermal line
broadening of low TR lines \citep[see, e.g.,][]{Erdelyi98,Chae98}.

One major issue that has hampered progress in our understanding of the
TR is that it has often been observed at inadequate spatial, temporal and
spectral resolution, or in isolation from the regions that impact its
dynamics such as the chromosphere and corona. The advent of IRIS \citep{DePontieu14} for the first time provides
spectra and images of the TR at high spatial
(0.33\arcsec), temporal ($\sim2$s) and spectral (3 km/s) resolution,
all of which are significant improvements over previous
instrumentation such as SUMER with resolutions of respectively,
2\arcsec, 10s and 8 km/s \citep{Wilhelm95}. In addition, IRIS provides
chromospheric spectra and images in the strong \ion{Mg}{2} h 
spectral line at the same time and location as the TR spectra. 

In this paper we focus on high-resolution observations of the
\ion{Si}{4} 1402.7\AA\ spectral line in active regions (AR), quiet Sun
(QS) and coronal hole (CH). We provide an overview of the datasets used in \S 2. In
\S 3 we describe how non-thermal line broadening and its correlation with
intensity is invariant to spatial and temporal summing, and how some
of the increased line broadening appears to be caused by chromospheric shocks. We
compare our observations with advanced numerical simulations that
include the effects of non-equilibrium ionization in the TR in \S 4. We finish with a discussion of the impact of our results in \S 5.

\begin{figure*}[!t]
\includegraphics[width=2\columnwidth]{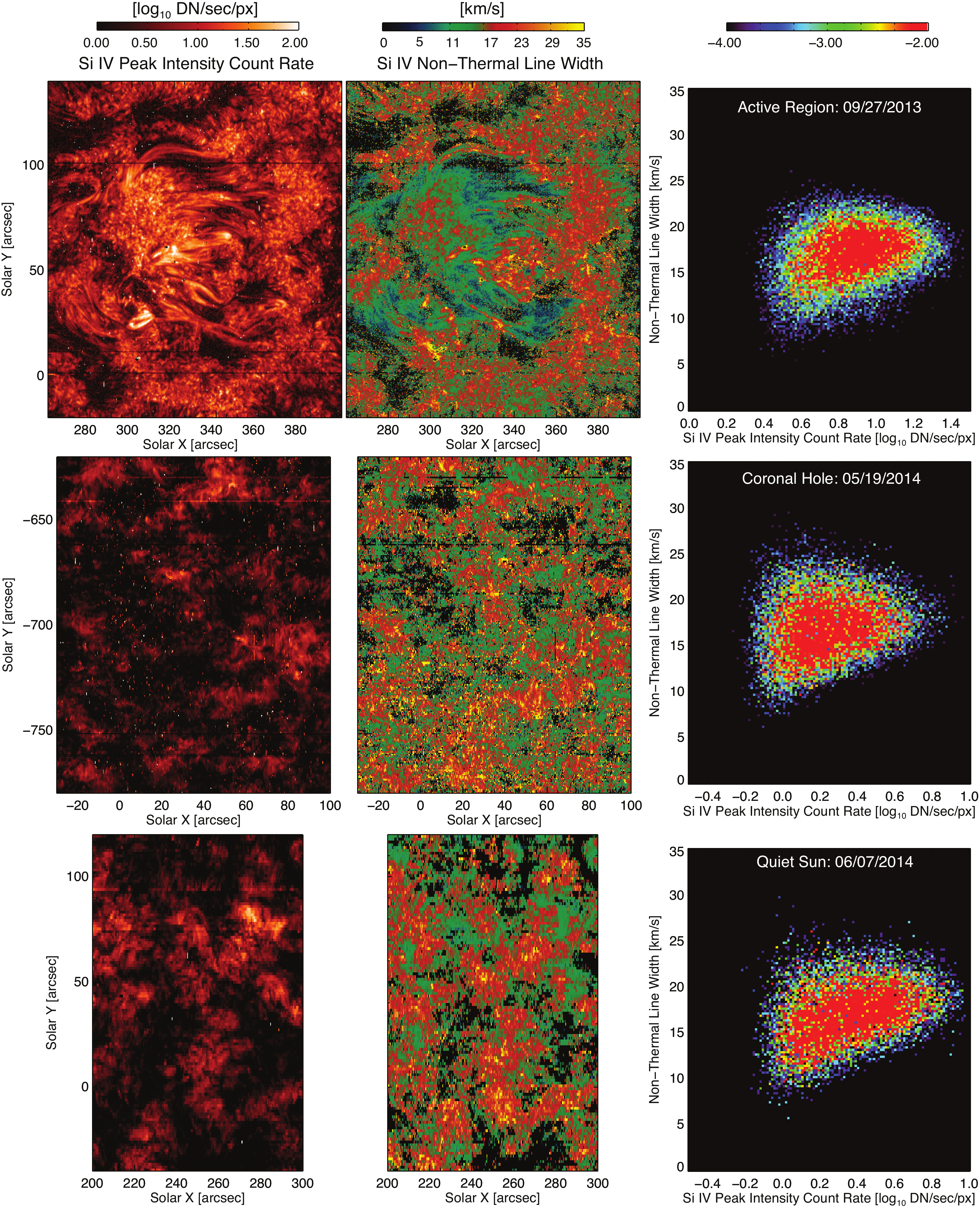}
\caption{Comparison of \ion{S}{4} 1403\AA{} peak line intensity and non-thermal widths for three different solar domains in rows from
top to bottom: active region, coronal hole, and quiet Sun for a single Gaussian fit to the observed IRIS line profiles. The left column of panels 
show the spatial variation of the (log$_{10}$) peak line intensity, while the central column shows the corresponding non-thermal line width 
and the right column shows the two-dimensional histogram of the former quantities.}
\label{fig1}
\end{figure*}

\section{Datasets}
\label{sec:datasets}

We use several different IRIS datasets: three raster scans and two
sit-and-stare sequences. The raster scans were taken on 2013-Sep-27 for
AR, 2014-June-7 for QS and 2013-May-19 for CH. The sit-and-stare
sequences (including solar rotation compensation) were both taken on
2013-Sep-10 with the AR sequence starting at 08:09 UT and the CH
sequence starting at 23:09 UT. All of the IRIS data were calibrated
to level 2, i.e., including dark current, flat-field and geometric
correction, as well as co-alignment between various channels. We also
corrected for the spectral drift associated with the spacecraft's
orbital velocity and the drift caused by
thermal variations during one orbit. The slit-jaw images
(SJI) were corrected for dark-current and flat-field, as well as
internal co-alignment drifts  \citep{DePontieu14}. 
We perform single Gaussian fits to spectral line profiles in locations
with sufficient signal-to-noise. From the fits we determine the peak
intensity $I_p$, the Doppler shift $v_D$ and the 1/e line width
$\sigma$. The latter is used to determine the non-thermal line
broadening $\sigma_{nt}$ with $\sigma_{nt} = \sqrt{\sigma^2 -
  \sigma_{th}^2 - \sigma_{inst}}$ where $\sigma_{th} = 6.86$ km/s
\citep[assuming a formation temperature of 80,000 K, see also][]{McIntosh08} and
$\sigma_{inst} = 3.9$ km/s \citep{DePontieu14}.  

\section{Observations}
\label{sec:observations}

\begin{figure*}[!t]
\includegraphics[width=2\columnwidth]{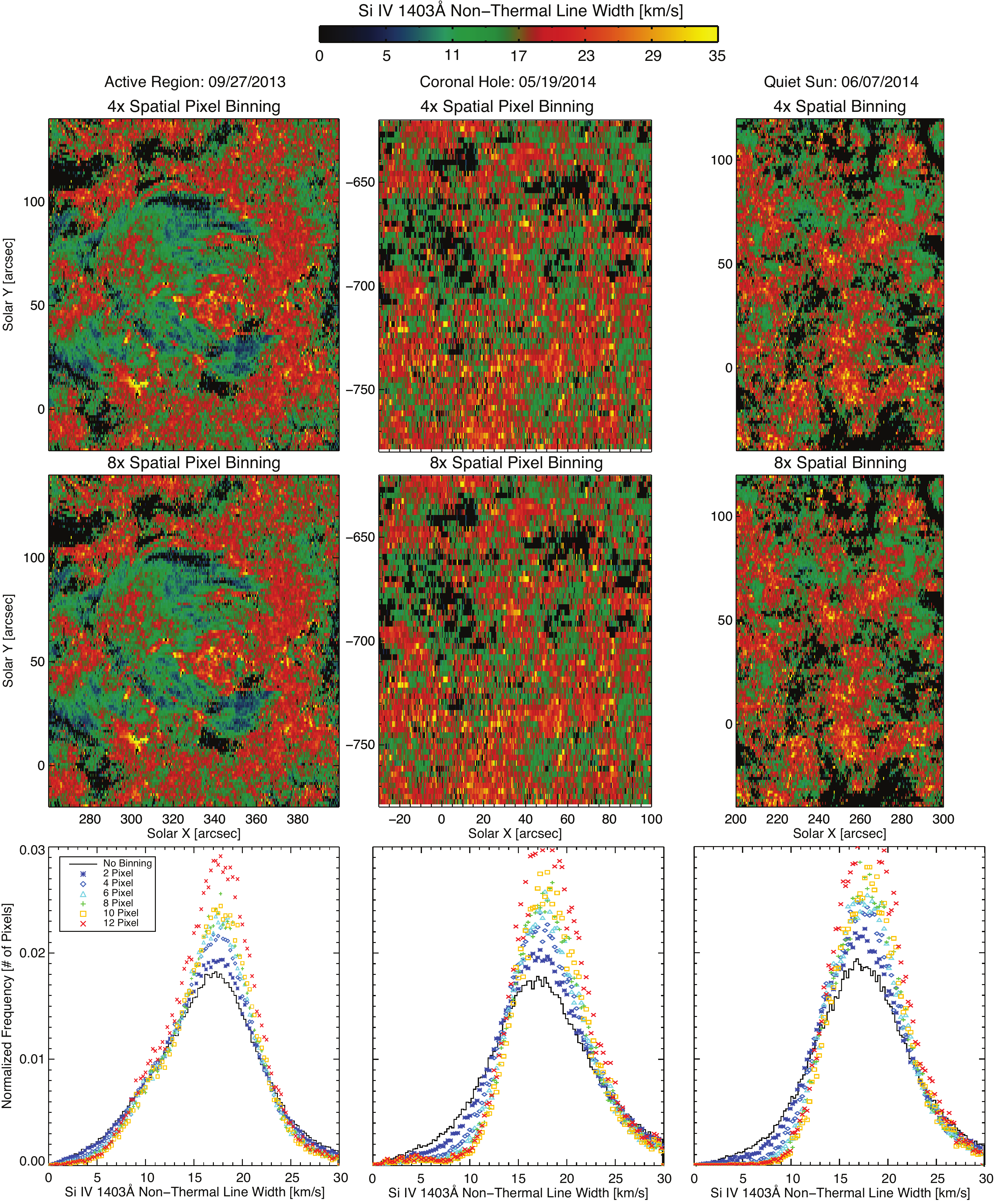}
\caption{Variations of the active region (left), coronal hole (center) and quiet sun (right) \ion{Si}{4} 1403\AA{} non-thermal line widths when 
different levels of spatial binning are applied to the \ion{Si}{4} line spectra prior to fitting with a single Gaussian profile (cf. the central panels of Fig.~\ref{fig1}).
The top row of panels shows the non-thermal line width when four spatial pixels are co-added in each case while the central row shows the same quantities
with eight spatial pixels co-added. The bottom row of panels shows the variation of the non-thermal line width at a range of spatial bindings from the native 
resolution of the instrument to twelve pixels as indicated in the panel legend (bottom left).}
\label{fig2}
\end{figure*}

IRIS observations of \ion{Si}{4} in AR, QS and CH show a variety
of bright features that include footpoints of loops (plage in AR,
network in QS) or the solar wind (network in CH) as well as less
bright regions, typically away from the strong flux concentrations
that underlie the bright network/plage regions. Comparison of
the left panels with the middle panels of Fig.~\ref{fig1} shows that the non-thermal 
line broadening is often enhanced in the brightest
regions. This correlation is not perfect on a pixel-to-pixel basis,
but clearly exists on larger scales. This is confirmed by the right panels
of Fig.~\ref{fig1} which show how the non-thermal line broadening is correlated with
the logarithm of the peak intensity of the line. This correlation was
previously found using spectra from lower resolution instruments \citep{Mariska1992}, but
our observations show that it does not change character at the very
high spatial and temporal resolution of IRIS.

Surprisingly, we find that the non-thermal line broadening is of order
20 km/s, even at the very high spatial resolution of 0.16\arcsec
(pixel size) by 0.33\arcsec (slit width) that
IRIS provides. More surprising is that the average
non-thermal line broadening observed with IRIS for AR, QS and CH 
are similar in value to measurements with lower resolution instruments
\citep[e.g.][]{Chae98}. This is confirmed when we spatially rebin our
observations to superpixels that are 2x2 original IRIS pixels and once
again perform a single Gaussian fit to the spectra. The
resulting histogram of non-thermal line broadening is essentially
identical to the histogram at IRIS native resolution (Fig.~\ref{fig2}). This invariance
to spatial summing occurs at 4x4 and 8x8 spatial resolution as well,
as shown in Fig.~\ref{fig2}. A similar invariance occurs for temporal
summing.

 At first sight it may seem surprising that the significant increase in spatial resolution of IRIS
does not lead to reduced non-thermal line broadening, since a map of
Doppler shifts indicates that there is a significant variation of the
line-of-sight velocity on subarcsecond spatial scales (not shown). However, the variation in line-of-sight velocities
(or macro-turbulence) on arcsecond scales is clearly not large enough
to cause significant additional broadening. Instead, the invariance to spatial and temporal summing indicates that the
processes responsible for the non-thermal line broadening either occur
on spatial scales that are smaller than the native IRIS resolution
(0.16\arcsec x 0.33\arcsec) and/or occur along the line-of-sight
(since \ion{Si}{4} 1402.7\AA\ is optically thin). 

\begin{figure*}[!t]
\includegraphics[width=\columnwidth]{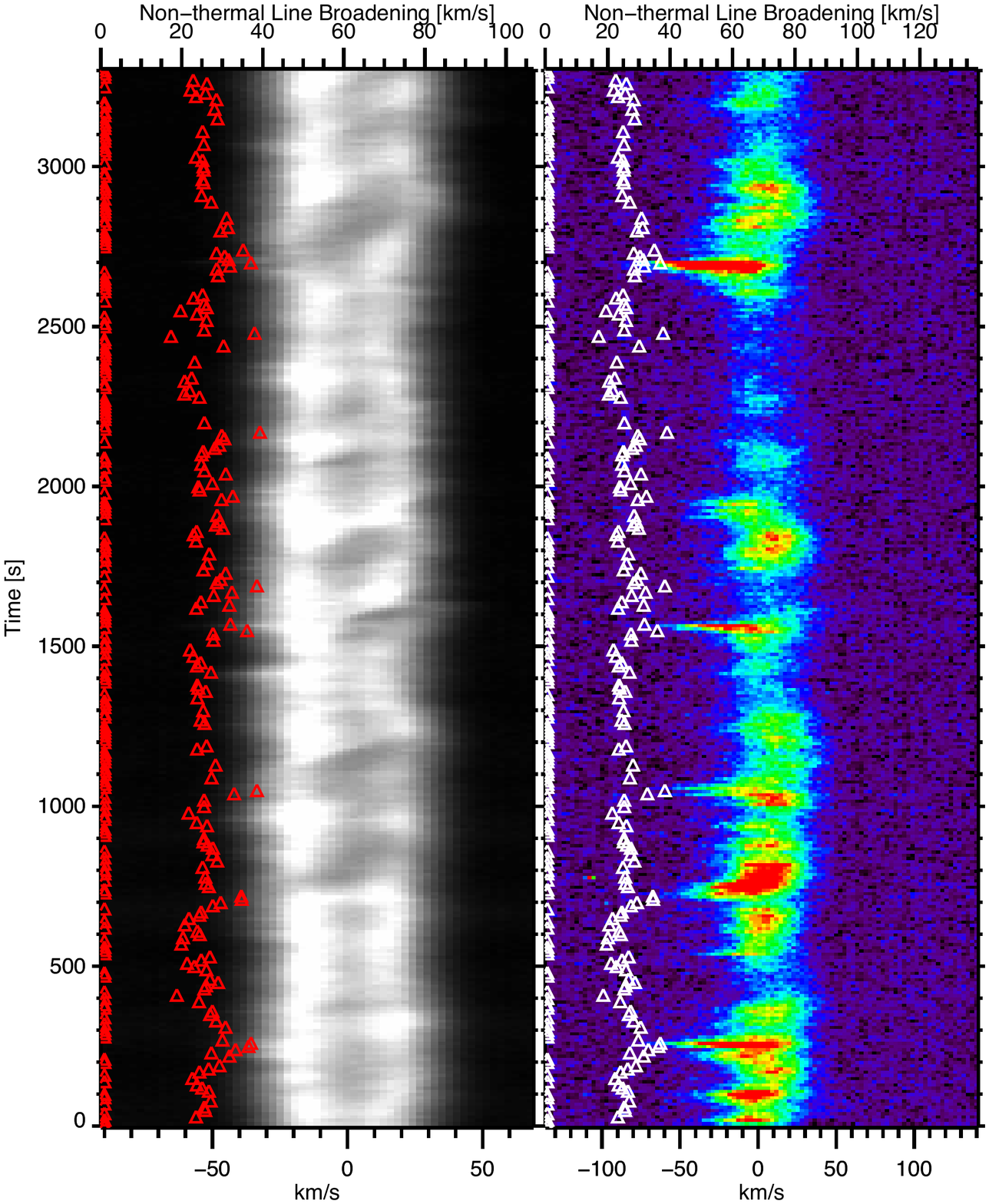}
\includegraphics[width=\columnwidth]{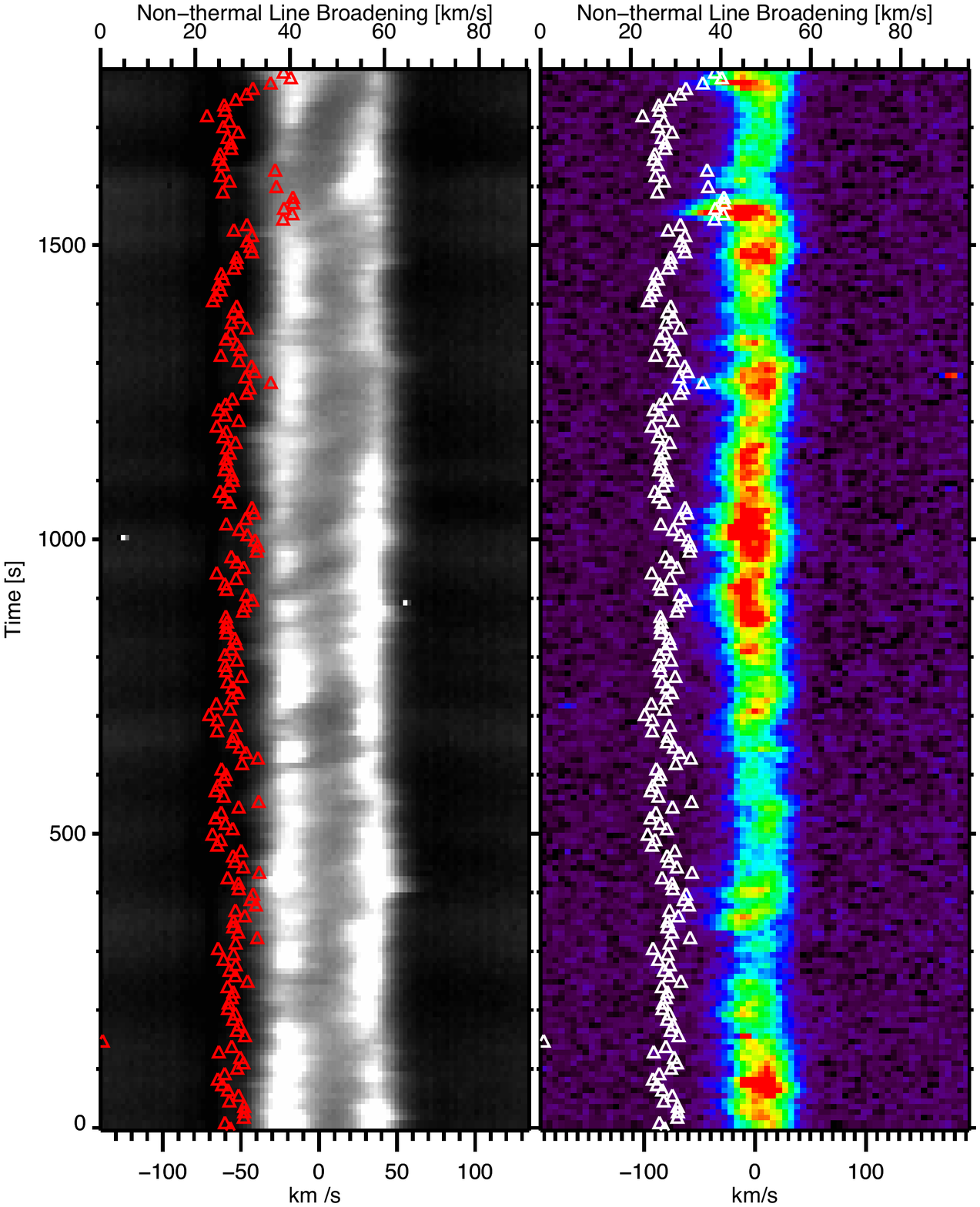}
\caption{Wavelength vs. time ($\lambda$-t) plots for locations in plage (a, b) and a network
  region in a coronal hole (c, d) for the \ion{Mg}{2} h 2803\AA\ line (a, c)
  and the \ion{Si}{4} 1402.7\AA\ line (b, d). Overplotted in red and white is
the evolution of the non-thermal line broadening as derived from a
single Gaussian fit to the \ion{Si}{4} line. The non-thermal line
broadening scale is shown on the top axis. The non-thermal line
broadening is set to 0 for profiles in which the signal is too weak
and/or the single Gaussian fit is poor. Notice the correlation
between shock occurrence (swing from red to blue) in the \ion{Mg}{2} h
line and the increased line broadening and brightness in \ion{Si}{4}.}
\label{fig3}
\end{figure*}

What are these processes, and can we find evidence for them in
our data? First, \citet{DePontieu14b} report on how the chromosphere and
TR are replete with twisting motions on spatial scales
down to the IRIS resolution. Their findings and previous
results indicate that twist
occurs on a variety of spatial scales, from macro-spicules and
tornadoes down to the 0.33 arcsec scales IRIS can observe, and that
this distribution of twist extends also to scales that are smaller than IRIS can
resolve. \citet{DePontieu14b} associate this twist with propagating
Alfv\'en waves with amplitudes of order 10-30 km/s and illustrate how
it becomes most easily visible when the line-of-sight is more perpendicular
to the direction of the magnetic field. In our data we find
support for such a scenario with many of the apparently low-lying
loops that connect the two plage regions in Fig.~\ref{fig1} showing
enhanced non-thermal line broadening. It is clear that significant
non-thermal line broadening would occur because of twist that is
part of the same distribution as \citet{DePontieu14b} report but on
smaller scales than IRIS can observe.

However, here we focus on another mechanism that
has been overlooked and appears to act when the line-of-sight is
more parallel to the magnetic field. In locations of increased
line broadening in plage/network regions, i.e., where the
magnetic field is typically more vertical, we find a correlation between increased line
broadening and the passage of magneto-acoustic shocks driven from the
chromosphere (Fig.~\ref{fig3}). These shocks are typically upward
propagating and have previously been reported as being ubiquitous in
the magnetic chromosphere
\citep[e.g.][]{Hansteen06,DePontieu07,Rouppe07,Langangen08a,Langangen08b,Vecchio09}. The
formation of these shocks is thought to
be caused by several mechanisms, e.g., leakage of
photospheric waves, convective motions or magnetic energy release in
the photosphere
\citep[e.g.][]{Heggland07,Martinez09, Heggland11}. The subsequent
upward propagation along a guiding magnetic field leads to slow-mode
magneto-acoustic shocks in a low plasma $\beta$ environment. The
shocks are recognizable as saw-tooth patterns in so-called
$\lambda-t$ (wavelength-time) plots of the \ion{Mg}{2} h line (Fig.~\ref{fig3}). The saw-tooth patterns show how the h3 spectral
feature repeatedly swings abruptly from the red wing to the blue wing
of the \ion{Mg}{2} h line as the shock propagates through the line
formation region of the h3 feature \citep[i.e., the top of the
chromosphere,][]{Leenaarts13a,Leenaarts13b,Pereira13}. Figure~\ref{fig3}
shows that these chromospheric shocks have a variable impact on the
\ion{Si}{4} spectral line profiles. Sometimes the shock signal is
clearly visible in the TR line as a rapid brightness excursion to the
blue followed by a linear shift with time to the red (e.g., panel B of
Fig.~\ref{fig3} around t=700-900s). At other times a shock signal is
present in the TR, but at much lower intensities (e.g., t=2100 s in
panel B of Fig.~\ref{fig3}). However, it is clear that in a plage
region (Fig.~\ref{fig3}a,b) the rapid excursions to the blue at the
time of a Mg II h shock (Fig.~\ref{fig3}a) are often brightest in
\ion{Si}{4}. In addition, the line width from a single Gaussian fit to
\ion{Si}{4} 1402.7\AA\ (white or red symbols, Fig.~\ref{fig3}a) often
significantly increases when the shock passage leads to a very rapid
switch from red to blueshifts, e.g., at $t=200, 700, 1050,1550,2700$
s. This suggests that the increased non-thermal line broadening in
these locations may be caused by the large range of velocities in the
pre- and post-shock environment, both of which are covered by the line
formation region. 
An alternative interpretation would be turbulence
caused by the shock passage, although it is unclear why the increased broadening would occur when the switch
from red to blueshifts happens.

The former scenario seems to hold for the CH network region 
although the lack of a strong guiding magnetic field (compared to
plage) leads to a less clear one-to-one correlation between shock
signatures in \ion{Mg}{2} h and \ion{Si}{4}. The magnetic
field in coronal holes is less strong and not as vertical as in plage,
which can easily lead to slight spatial displacements between the
chromospheric shock and the TR counterpart. Nevertheless, there is
still a correlation between strong chromospheric velocity excursions
and brightenings and associated non-thermal line broadening in the TR.

Clearly, the influence of magneto-acoustic shocks and their impact on
non-thermal line broadening could provide a straightforward explanation for
the observed invariance to spatio-temporal summing of the
broadening. It could also explain why there is a correlation between
non-thermal broadening and the intensity, with shock heating causing
strong brightening, as well as non-thermal line broadening.

We note that the non-thermal line
broadening is also increased during times when Mg II h is not
switching from strong red- to blueshifts.

\section{Numerical Simulations} 
\label{sec:results}

While these observations provide a possible reason for the
observed invariance and correlations, they are also puzzling. In
particular, one would not expect the line formation region
of \ion{Si}{4} 1402.7\AA\ to cover both the pre- and post-shock
environment given that under ionization equilibrium conditions the line 
is expected to from mostly in a narrow temperature range around
80,000$\pm 20,000$K. Numerical simulations of a computational
domain ranging from the top of the convection zone into the corona help shed
light on this. We performed a 2.5D radiative MHD simulation using
the Bifrost code \citep{Gudiksen11} in which the main free parameter
is the magnetic field configuration. Bifrost includes thermal
conduction and optically thin radiative losses in the corona, as well
as radiative losses in the optically thick chromosphere
and photosphere. The initial field configuration was chosen
to be representative of a coronal hole (average unsigned flux of 5
Mx~cm$^{-2}$). We calculated the Si ionization in two different ways: 1. ionization equilibrium, 2. 
non-equilibrium, i.e., time-dependent, ionization of Si \citep[using
the approach of][]{Olluri13}. We calculate \ion{Si}{4} 1402.7\AA\ emission using
CHIANTI \citep{Landi2012} and optically thick \ion{Mg}{2} h 2803\AA\ spectral line
profiles using RH 1.5D \citep{Pereira2014}.

\begin{figure*}[!t]
\includegraphics[width=2\columnwidth]{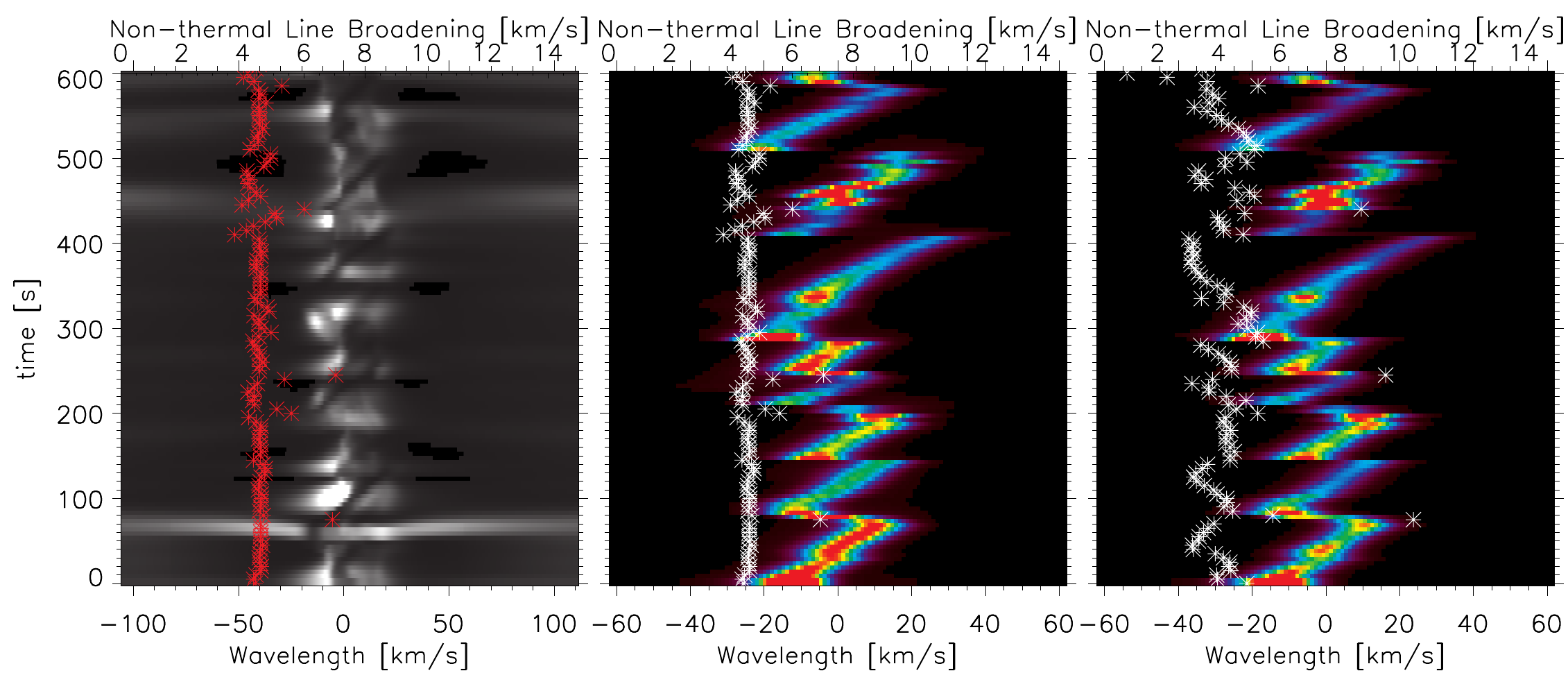}
\caption{$\lambda$ t plots for the \ion{Mg}{2} h 2803\AA\ line (a)
  and the \ion{Si}{4} 1402.7\AA\ line (b, c) from numerical
  simulations of a unipolar environment assuming ionization equilibrium (a,
  b) and non-equilibrium ionization for Si$^{3+}$ (c). 
  Overplotted in red or white are
  the evolution of the non-thermal line broadening as derived from a
  single Gaussian fit to the \ion{Si}{4} line. Notice the correlation
  between shock occurrence (swing from red to blue) in the \ion{Mg}{2} h
  line and the increased line broadening and brightness in \ion{Si}{4}.}
\label{fig4}
\end{figure*}

\begin{figure}[!t]
\includegraphics[width=\columnwidth]{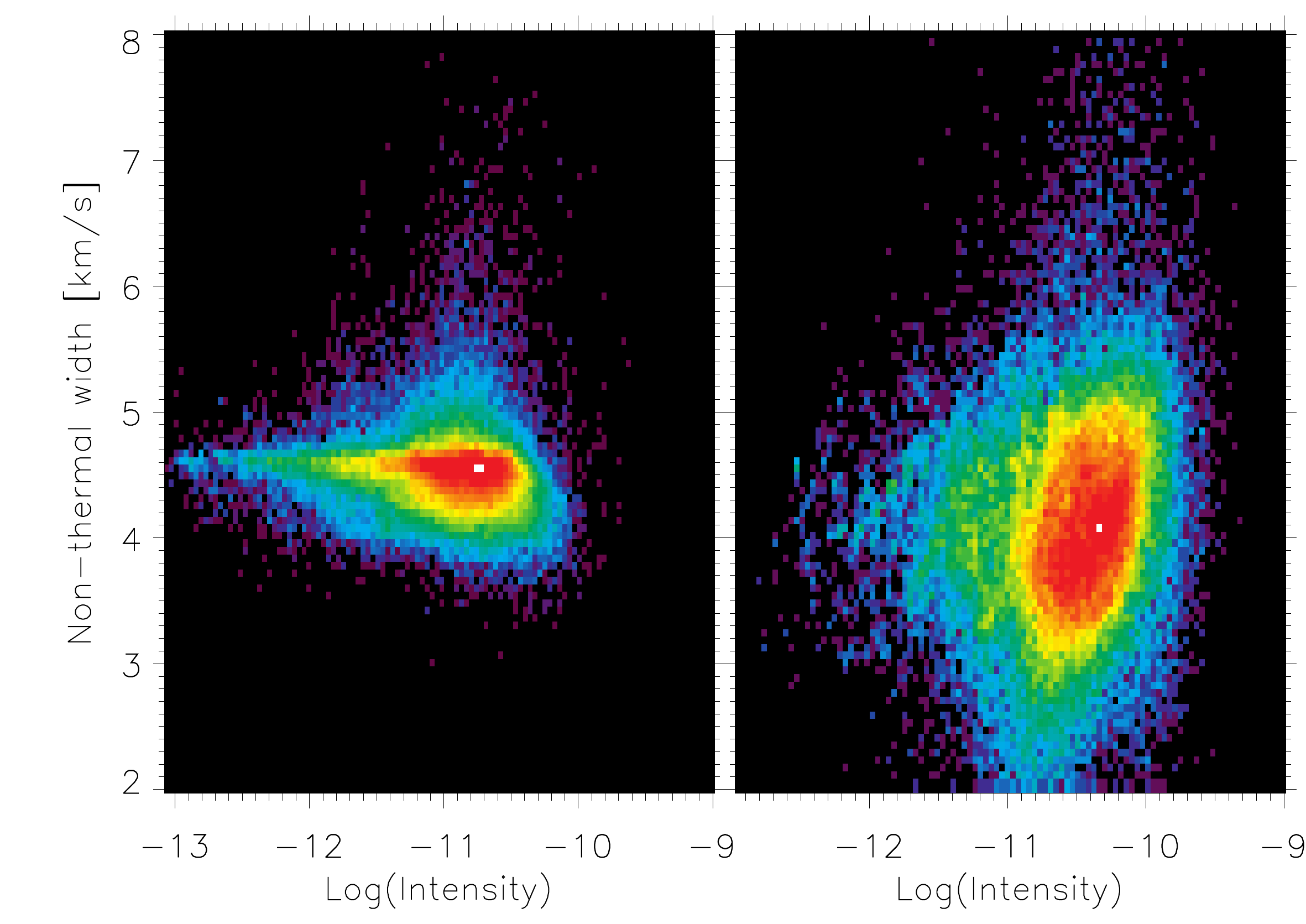}
\caption{Probability density function between logarithm of the
  \ion{Si}{4} 1402.7\AA\ intensity and its non-thermal line
  broadening, for simulations with ionization equilibrium (left) and 
  including non-equilibrium ionization (right). Note the lack of correlation
  in the ionization equilibrium simulations, and the appearance of a correlation for non-equilibrium ionization.}
\label{fig5}
\end{figure}

We find that both results show evidence of
magneto-acoustic shocks that propagate from the chromosphere into the
TR. As shown in Fig.~\ref{fig4} the simulated shocks lead to increased non-thermal line broadening in the TR at the time
of the shock passage. However, this is more evident in the simulation
assuming non-equilibrium ionization: the broadening is
  increased by 2-10 km/s during shock passage, and by several km/s
  throughout the simulation. 
More importantly, the ionization equilibrium simulation does not
reproduce the correlation (Fig.~\ref{fig5}) between non-thermal line
broadening and the logarithm of the intensity. That correlation only
appears when non-equilibrium ionization is included in the simulations, as
shown in Fig.~\ref{fig5}. 

This is because the non-equilibrium ionization leads
to the presence of Si$^{3+}$ ions over a much wider range of
temperatures than under ionization equilibrium \cite[see,
e.g.][]{Olluri13}: many locations show  Si$^{3+}$ ionization fractions 
 above 20\% over a temperature range from $10,000$ to
  $200,000$ K, with most of the \ion{Si}{4} emission resulting from
  plasma at temperatures between 20,000 and 200,000 K. This means that the line formation region more
easily captures both the pre- and post-shock environment. This
naturally leads to a larger range of velocities along the line-of-sight of
the optically thin \ion{Si}{4} lines, and thus non-thermal line
broadening. This sensitivity to the shocks provides a natural
explanation for the correlation between intensity and non-thermal line broadening.

The models are numerical experiments. While
they reproduce aspects of our observations, they
are currently unable to reproduce the large non-thermal line
broadening that is observed. This is likely because the numerical
resistivity leads to energy deposition on larger spatial scales
than in the solar atmosphere, leading to less violent shocks and other events.

\section{Discussion \& Conclusion}
\label{sec:discussion}

Our IRIS observations and numerical simulations suggest that the observed invariance
to spatio-temporal resolution of the non-thermal line broadening of
\ion{Si}{4} 1402.7\AA, and its correlation with the intensity may
have two causes. First, in regions where the line-of-sight is
more perpendicular to the local magnetic field the presence of twisting
motions or other turbulent motions on sub-resolution scales likely plays a role in both effects. We find
direct evidence that in regions where the line-of-sight is more parallel
with the magnetic field (e.g., network or plage regions) the presence
of magneto-acoustic shocks along the line-of-sight can lead to
enhanced non-thermal line broadening, as well as its correlation with
the intensity. This sheds some doubt on previous studies which have
emphasized or hypothesized about how the correlation between intensity
and non-thermal broadening is indicative of or gives clues about the
coronal heating mechanism (nanoflares) or turbulence. Our results
suggest that this may not be the case and that instead the tight
connection between chromospheric and TR dynamics plays
an important role in the observed line broadening in the TR.

The correlation with slow-mode magneto-acoustic shock waves also
provides a natural explanation for why correlations between parameters
of TR lines with increasing temperature differences
become worse. The effects of the shocks at high temperatures are
diminished as thermal conduction starts to smooth out the strong
gradients associated with shocks.

The center-to-limb variation of the non-thermal line width is small,
if present at all \citep{Peter99}. While the effect of the shocks
(mostly ${||}B$) should decrease towards the limb, the effect of the
twisting motions (mostly ${\perp}B$) should increase towards the
limb. How these two effects combine to lead to a roughly constant
center-to-limb variation remains to be investigated in detail, but is
likely similar to the results of \citet{McIntosh08} who investigated
the trade-off between field-aligned shock-driven jets and transverse 
wave motions of the same size.

Our comparison of the observations with numerical modeling suggests
that non-equilibrium ionization may play an important role in the
diagnostics, dynamics and energetics of the TR, as
already suggested by advanced simulations \citep{Olluri13}.
The fact that non-equilibrium ionization leads to TR
emission that originates over a larger temperature/height range, 
implies that, in the presence of shocks with strong velocity
gradients, some of the non-thermal line broadening observed is caused
by relatively mundane slow-mode magneto-acoustic shocks. Such shocks
are plentiful in the chromosphere and it would be very surprising if
they did not affect the TR in some fashion. Other
effects that impact the chromosphere/transition region, such as
ambipolar diffusion caused by the interactions between ions and
neutrals, may also play a role in these types of diagnostics.

Several unresolved issues remain. While shocks clearly play a role in
causing non-thermal line broadening, especially at the time of impact
of the shock in the TR, the non-thermal line broadening
is significant even at times when the chromospheric velocity does not rapidly
shift from red to blueshifts. Perhaps the increased non-thermal line
broadening at other times is caused by the strong velocity gradients in the pre-shock or
post-shock plasma combined with non-equilibrium ionization and other effects that
lead to TR emission over a wide range of heights? Alternatively, the
simulations do not contain type 2 spicules which, based on their
observational characteristics, likely show strong flows and strong
gradients and could thus also contribute to non-thermal line broadening.

It also remains unclear how the pervasive redshifts in the
TR fit into the picture we have sketched in the
above. Further numerical and observational studies will be required to
settle that issue.



	
	
       
       
      

\acknowledgments
IRIS is a NASA Small Explorer developed and operated by LMSAL with
mission operations executed at NASA Ames Research center and major
contributions to downlink communications funded by the Norwegian Space
Center through an ESA PRODEX contract. This work is supported by NASA
contract NNG09FA40C (IRIS) and has benefited from
discussions at the International Space Science Institute (ISSI)
meeting on “Heating of the magnetized chromosphere”. The simulations were run on Pleiades through the computing project s1061 from the High End Computing (HEC) division of NASA.


\begin{thebibliography}{28}
\expandafter\ifx\csname natexlab\endcsname\relax\def\natexlab#1{#1}\fi

\bibitem[{{Chae} {et~al.}(1998){Chae}, {Sch{\"u}hle}, \& {Lemaire}}]{Chae98}
{Chae}, J., {Sch{\"u}hle}, U., \& {Lemaire}, P. 1998, \apj, 505, 957

\bibitem[{{De Pontieu} {et~al.}(2007){De Pontieu}, {Hansteen}, {Rouppe van der
  Voort}, {van Noort}, \& {Carlsson}}]{DePontieu07}
{De Pontieu}, B., {Hansteen}, V.~H., {Rouppe van der Voort}, L., {van Noort},
  M., \& {Carlsson}, M. 2007, \apj, 655, 624

\bibitem[{{De Pontieu} {et~al.}(2014{\natexlab{a}}){De Pontieu}, {Rouppe van
  der Voort}, {McIntosh}, {Pereira}, {Carlsson}, {Hansteen}, {Skogsrud},
  {Lemen}, {Title}, {Boerner}, {Hurlburt}, {Tarbell}, {Wuelser}, {De Luca},
  {Golub}, {McKillop}, {Reeves}, {Saar}, {Testa}, {Tian}, {Kankelborg},
  {Jaeggli}, {Kleint}, \& {Martinez-Sykora}}]{DePontieu14b}
{De Pontieu}, B., {Rouppe van der Voort}, L., {McIntosh}, S.~W., {et~al.}
  2014{\natexlab{a}}, Science, 346, D315

\bibitem[{{De Pontieu} {et~al.}(2014{\natexlab{b}}){De Pontieu}, {Title},
  {Lemen}, {Kushner}, {Akin}, {Allard}, {Berger}, {Boerner}, {Cheung}, {Chou},
  {Drake}, {Duncan}, {Freeland}, {Heyman}, {Hoffman}, {Hurlburt}, {Lindgren},
  {Mathur}, {Rehse}, {Sabolish}, {Seguin}, {Schrijver}, {Tarbell},
  {W{\"u}lser}, {Wolfson}, {Yanari}, {Mudge}, {Nguyen-Phuc}, {Timmons}, {van
  Bezooijen}, {Weingrod}, {Brookner}, {Butcher}, {Dougherty}, {Eder},
  {Knagenhjelm}, {Larsen}, {Mansir}, {Phan}, {Boyle}, {Cheimets}, {DeLuca},
  {Golub}, {Gates}, {Hertz}, {McKillop}, {Park}, {Perry}, {Podgorski},
  {Reeves}, {Saar}, {Testa}, {Tian}, {Weber}, {Dunn}, {Eccles}, {Jaeggli},
  {Kankelborg}, {Mashburn}, {Pust}, {Springer}, {Carvalho}, {Kleint}, {Marmie},
  {Mazmanian}, {Pereira}, {Sawyer}, {Strong}, {Worden}, {Carlsson}, {Hansteen},
  {Leenaarts}, {Wiesmann}, {Aloise}, {Chu}, {Bush}, {Scherrer}, {Brekke},
  {Martinez-Sykora}, {Lites}, {McIntosh}, {Uitenbroek}, {Okamoto}, {Gummin},
  {Auker}, {Jerram}, {Pool}, \& {Waltham}}]{DePontieu14}
{De Pontieu}, B., {Title}, A.~M., {Lemen}, J.~R., {et~al.} 2014{\natexlab{b}},
  \solphys, 289, 2733

\bibitem[{{Erdelyi} {et~al.}(1998){Erdelyi}, {Doyle}, {Perez}, \&
  {Wilhelm}}]{Erdelyi98}
{Erdelyi}, R., {Doyle}, J.~G., {Perez}, M.~E., \& {Wilhelm}, K. 1998, \aap,
  337, 287

\bibitem[{{Gudiksen} {et~al.}(2011){Gudiksen}, {Carlsson}, {Hansteen}, {Hayek},
  {Leenaarts}, \& {Mart{\'{\i}}nez-Sykora}}]{Gudiksen11}
{Gudiksen}, B.~V., {Carlsson}, M., {Hansteen}, V.~H., {et~al.} 2011, \aap, 531,
  A154

\bibitem[{{Hansteen} {et~al.}(2014){Hansteen}, {De Pontieu}, {Carlsson}, \&
  {Lemen}}]{Hansteen14}
{Hansteen}, V.~H., {De Pontieu}, B., {Carlsson}, M., \& {Lemen}, J. 2014,
  Science, 1, 1

\bibitem[{{Hansteen} {et~al.}(2006){Hansteen}, {De Pontieu}, {Rouppe van der
  Voort}, {van Noort}, \& {Carlsson}}]{Hansteen06}
{Hansteen}, V.~H., {De Pontieu}, B., {Rouppe van der Voort}, L., {van Noort},
  M., \& {Carlsson}, M. 2006, \apjl, 647, L73

\bibitem[{{Hansteen} {et~al.}(2010){Hansteen}, {Hara}, {De Pontieu}, \&
  {Carlsson}}]{Hansteen10}
{Hansteen}, V.~H., {Hara}, H., {De Pontieu}, B., \& {Carlsson}, M. 2010, \apj,
  718, 1070

\bibitem[{{Heggland} {et~al.}(2007){Heggland}, {De Pontieu}, \&
  {Hansteen}}]{Heggland07}
{Heggland}, L., {De Pontieu}, B., \& {Hansteen}, V.~H. 2007, \apj, 666, 1277

\bibitem[{{Heggland} {et~al.}(2011){Heggland}, {Hansteen}, {De Pontieu}, \&
  {Carlsson}}]{Heggland11}
{Heggland}, L., {Hansteen}, V.~H., {De Pontieu}, B., \& {Carlsson}, M. 2011,
  \apj, 743, 142

\bibitem[{{Landi} {et~al.}(2012){Landi}, {Del Zanna}, {Young}, {Dere}, \&
  {Mason}}]{Landi2012}
{Landi}, E., {Del Zanna}, G., {Young}, P.~R., {Dere}, K.~P., \& {Mason}, H.~E.
  2012, \apj, 744, 99

\bibitem[{{Langangen} {et~al.}(2008{\natexlab{a}}){Langangen}, {Carlsson},
  {Rouppe van der Voort}, {Hansteen}, \& {De Pontieu}}]{Langangen08b}
{Langangen}, {\O}., {Carlsson}, M., {Rouppe van der Voort}, L., {Hansteen}, V.,
  \& {De Pontieu}, B. 2008{\natexlab{a}}, \apj, 673, 1194

\bibitem[{{Langangen} {et~al.}(2008{\natexlab{b}}){Langangen}, {Rouppe van der
  Voort}, \& {Lin}}]{Langangen08a}
{Langangen}, {\O}., {Rouppe van der Voort}, L., \& {Lin}, Y.
  2008{\natexlab{b}}, \apj, 673, 1201

\bibitem[{{Leenaarts} {et~al.}(2013{\natexlab{a}}){Leenaarts}, {Pereira},
  {Carlsson}, {Uitenbroek}, \& {De Pontieu}}]{Leenaarts13a}
{Leenaarts}, J., {Pereira}, T.~M.~D., {Carlsson}, M., {Uitenbroek}, H., \& {De
  Pontieu}, B. 2013{\natexlab{a}}, \apj, 772, 89

\bibitem[{{Leenaarts} {et~al.}(2013{\natexlab{b}}){Leenaarts}, {Pereira},
  {Carlsson}, {Uitenbroek}, \& {De Pontieu}}]{Leenaarts13b}
{Leenaarts}, J., {Pereira}, T.~M.~D., {Carlsson}, M., {Uitenbroek}, H., \& {De
  Pontieu}, B. 2013{\natexlab{b}}, \apj, 772, 90

\bibitem[{{Mariska}(1992)}]{Mariska1992}
{Mariska}, J.~T. 1992, {The solar transition region}

\bibitem[{{Mart{\'{\i}}nez-Sykora} {et~al.}(2009){Mart{\'{\i}}nez-Sykora},
  {Hansteen}, {De Pontieu}, \& {Carlsson}}]{Martinez09}
{Mart{\'{\i}}nez-Sykora}, J., {Hansteen}, V., {De Pontieu}, B., \& {Carlsson},
  M. 2009, \apj, 701, 1569

\bibitem[{{McIntosh} {et~al.}(2008){McIntosh}, {De Pontieu}, \&
  {Tarbell}}]{McIntosh08}
{McIntosh}, S.~W., {De Pontieu}, B., \& {Tarbell}, T.~D. 2008, \apjl, 673, L219

\bibitem[{{Olluri} {et~al.}(2013){Olluri}, {Gudiksen}, \&
  {Hansteen}}]{Olluri13}
{Olluri}, K., {Gudiksen}, B.~V., \& {Hansteen}, V.~H. 2013, \apj, 767, 43

\bibitem[{{Pereira} {et~al.}(2013){Pereira}, {Leenaarts}, {De Pontieu},
  {Carlsson}, \& {Uitenbroek}}]{Pereira13}
{Pereira}, T.~M.~D., {Leenaarts}, J., {De Pontieu}, B., {Carlsson}, M., \&
  {Uitenbroek}, H. 2013, \apj, 778, 143

\bibitem[{{Pereira} \& {Uitenbroek}(2014)}]{Pereira2014}
{Pereira}, T.~M.~D. \& {Uitenbroek}, H. 2014, ArXiv e-prints

\bibitem[{{Peter}(1999)}]{Peter99b}
{Peter}, H. 1999, \apj, 516, 490

\bibitem[{{Peter} {et~al.}(2004){Peter}, {Gudiksen}, \& {Nordlund}}]{Peter2004}
{Peter}, H., {Gudiksen}, B.~V., \& {Nordlund}, {\AA}. 2004, \apjl, 617, L85

\bibitem[{{Peter} \& {Judge}(1999)}]{Peter99}
{Peter}, H. \& {Judge}, P.~G. 1999, \apj, 522, 1148

\bibitem[{{Rouppe van der Voort} {et~al.}(2007){Rouppe van der Voort}, {De
  Pontieu}, {Hansteen}, {Carlsson}, \& {van Noort}}]{Rouppe07}
{Rouppe van der Voort}, L.~H.~M., {De Pontieu}, B., {Hansteen}, V.~H.,
  {Carlsson}, M., \& {van Noort}, M. 2007, \apjl, 660, L169

\bibitem[{{Vecchio} {et~al.}(2009){Vecchio}, {Cauzzi}, \&
  {Reardon}}]{Vecchio09}
{Vecchio}, A., {Cauzzi}, G., \& {Reardon}, K.~P. 2009, \aap, 494, 269

\bibitem[{{Wilhelm} {et~al.}(1995){Wilhelm}, {Curdt}, {Marsch}, {Sch{\"u}hle},
  {Lemaire}, {Gabriel}, {Vial}, {Grewing}, {Huber}, {Jordan}, {Poland},
  {Thomas}, {K{\"u}hne}, {Timothy}, {Hassler}, \& {Siegmund}}]{Wilhelm95}
{Wilhelm}, K., {Curdt}, W., {Marsch}, E., {et~al.} 1995, \solphys, 162, 189

\end{thebibliography}

\end{document}